\documentclass[aps,pra,a4paper,preprint,nofootinbib]{revtex4-1}

\usepackage{amsfonts,amsmath,amssymb,bm,tensor}
\usepackage{comment}
\usepackage{ifpdf}
\usepackage{slashed}
\usepackage{braket}
\usepackage[mathscr]{eucal}
\usepackage[utf8]{inputenc}
\usepackage{cancel}

\ifpdf
\usepackage{graphicx}     
\usepackage[bookmarksopen,colorlinks=true,linkcolor=bblue,citecolor=bblue,urlcolor=ppink]{hyperref}
\else     
\usepackage[dvipdfmx]{graphicx}     
\usepackage[dvipdfmx,bookmarksopen,colorlinks=true,linkcolor=bblue,citecolor=ppink,urlcolor=darkred]{hyperref}
\fi

\usepackage{color}
\definecolor{darkred}{rgb}{0.6,0,0}
\definecolor{darkgreen}{rgb}{0.992447,0.623778,0.034597}
\definecolor{ppink}{rgb}{1,0.4,0.4}
\definecolor{bblue}{rgb}{0.284602,0.317763,0.963947}
\definecolor{mygreen}{rgb}{0,0.7,0}
\definecolor{myred}{rgb}{1,0.3,0.4}
\definecolor{myblue}{rgb}{0.2,0.3,1}

	\newcommand{\eV}{\mathrm{eV}}
	\newcommand{\keV}{\mathrm{keV}}
	
	\newcommand{\GeV}{\mathrm{GeV}}

\newcommand{\tx}{\text}

\newcommand{\qcq}{\quad,\quad}

\newcommand{\df}{\text{d}}
\newcommand{\p}{\partial}

\begin{document}


\title{Revisiting CMB constraints on dark matter annihilation}

\author{Masahiro Kawasaki}
\affiliation{ICRR, University of Tokyo, Kashiwa, 277-8582, Japan}
\affiliation{Kavli IPMU (WPI), UTIAS, University of Tokyo, Kashiwa, 277-8583, Japan}
\author{Hiromasa Nakatsuka}
\affiliation{ICRR, University of Tokyo, Kashiwa, 277-8582, Japan}
\author{Kazunori Nakayama}
\affiliation{Department of Physics, Faculty of Science, The University of Tokyo, Bunkyo-ku, Tokyo 113-0033, Japan}
\author{Toyokazu Sekiguchi}
\affiliation{Theory Center, IPNS, KEK, Tsukuba, Ibaraki 305-0801, Japan}

\begin{abstract}
	\noindent

   The precision measurements of the cosmic microwave background power spectrum put a strong constraint on the dark matter annihilation cross section since the electromagnetic energy injection by the dark matter annihilation affects the ionization history of the universe.
    In this paper, we update our previous simulation code for calculating the ionization history with the effect of dark matter annihilation by including Helium interactions and improving the precision of calculations. 
    We give an updated constraint on the annihilation cross section and mass of dark matter using the modified {RECFAST} code with the Planck 2018 datasets.

\end{abstract}

\date{\today}
\maketitle
\tableofcontents

\section{Introduction}
\label{sec_Introduction}

The existence of dark matter (DM), contributing to $24$\% of the present energy density of the universe, is one of the most important mysteries in cosmology and particle physics.
DM is likely composed of particles beyond the standard model of particle physics, and many candidates have been proposed so far.
Among them, WIMPs (weakly interacting particles) have been most intensively studied and searched for directly and indirectly.

The so-called { indirect searches} look for the DM annihilation signals. Several observations can put constraints on the DM annihilation cross section~$\braket{\sigma v}$: CMB (Cosmic microwave background)~\cite{Padmanabhan:2005es,Mapelli:2006ej,Zhang:2006fr,Kanzaki:2008qb,Kanzaki:2009hf,Slatyer:2009yq,Slatyer:2012yq,Galli:2013dna,Slatyer:2015jla,Slatyer:2015kla,Liu:2016cnk,Acharya:2019uba,Cang:2020exa}, $\gamma$-rays~\cite{Ackermann:2015zua,Ahnen:2016qkx}, cosmic rays~\cite{Aguilar:2013qda,Adriani:2008zr,Aharonian:2009ah,ANTARES:2019svn,Gelmini:2015zpa} and light elements synthesized in BBN (Big Bang Nucleosynthesis) ~\cite{Reno:1987qw,Jedamzik:2004ip,Hisano:2009rc,Hisano:2011dc,Kawasaki:2015yya,Depta:2019lbe}.
For the DM mass about $10$--$1000$\,GeV, the $\gamma$-ray observations of dwarf galaxies may give the most stringent constraint~\cite{VERITAS:2010meb,HESS:2014zqa,Ackermann:2015zua,Ahnen:2016qkx}.
However, the constraint may contain sizable astronomical uncertainties because it is sensitive to the DM profiles in dwarf galaxies which are difficult to determine precisely. 
On the other hand, CMB can give a robust constraint, which does not suffer from such astronomical uncertainties, since the dominant effect of DM annihilation on the CMB comes from the epoch around the recombination when the DM spatial distribution is almost homogeneous.
In the early universe, the DM annihilation into the standard model particles injects energy into the background plasma and it changes the ionization history of the universe around/after the recombination epoch ($z<3000$ in terms of the redshift)~\cite{Adams:1998nr,Chen:2003gz,Poulin:2016anj}.
This then affects the CMB angular power spectrum.
The Planck observation reaches the $\mathcal O(1)\%$ precision in determining the various cosmological parameters, which strongly constrains the energy injection from the DM annihilation in the early universe.

Let us focus on the CMB constraint on DM annihilation.
In order to obtain a reliable constraint, we have to estimate the ionization fractions of hydrogen and helium taking all relevant radiative processes into account.
In the literature, the efficiency factor $f_\tx{eff}(z)$ is often used for parameterizing the energy injection of DM annihilation.
$f_\tx{eff}(z)$ is the ratio of absorbed energy to the injected one at a given redshift~\cite{Aghanim:2018eyx} and it depends on the interaction between the background plasma and the high energy particles injected by the DM annihilation.
The high-energy particles lose their energies by causing electromagnetic shower, in which injected particles produce a lot of scattered particles with energy large enough to ionize  hydrogen and helium.
The electromagnetic interactions that take place in the showers are usually faster than the cosmic expansion.
However, the cosmic plasma is transparent against photons with high energy ($10^8$~eV -$10^{10}$~eV at $z\sim 10^3$) and they lose their energy slowly in a longer time-scale than the Hubble time~\cite{Chen:2003gz,Slatyer:2009yq}, for which $f_\tx{eff}(z)$ cannot describe the energy injection correctly.
Thus we need to solve the shower process over the whole energy and time ranges, by piling up every interaction from low to high energy and from low to high redshift~\cite{Kanzaki:2008qb,Kanzaki:2009hf,Slatyer:2009yq}.
The high-energy particles finally lose their energy by heating the plasma, ionizing hydrogen and helium, and the cosmic expansion.
The modified ionization history changes the CMB power spectrum.
Using the current CMB data, we can constrain the DM parameter space $(m_\tx{DM},\braket{\sigma v})$ where $m_\text{DM}$ is the DM mass.
We use the Markov Chain Monte Carlo methods (MCMC) to estimate the allowed parameter space.

Particle production and scattering in electromagnetic showers have been investigated by several works.
Among them, Refs.~\cite{Galli:2013dna,Slatyer:2015jla,Slatyer:2015kla} separated the shower process into the low and high energy parts, where the Monte Carlo method is used for cooling of electrons through atomic processes with energy smaller than $3\,$keV and the recursive method is used with simplified interactions with energy larger than $3\,$keV.
In the high energy regime, for example, the absorbed energy by photoionization is neglected and the collisional ionization on helium is treated as that of hydrogen.
They calculated the shower process with very fine time steps $\Delta (\log z)=10^{-3}$ and made the table of the ratio of the injected  energy to that used for various processes (e.g. ionization) with 40 log-spaced bins for energy between 1 keV and 10 TeV and 63 log-spaced bins for redshift from $1+z=10$ to 3000.
On the other hand, in our previous works~\cite{Kanzaki:2008qb,Kanzaki:2009hf,Kawasaki:2015peu}, we efficiently calculated the shower process over the whole energy range by recursive method with the ``discrete'' and ``continuous'' interactions, where the scattered particle loses larger (smaller) energy than the size of the energy bin $\Delta E$ in discrete (continuous) interaction.
We can consistently calculate shower process in both high and low energy regions in this method without unnecessary assumptions.
Some other works \cite{Acharya:2018iwh,Acharya:2019uba} also adopted this method and calculated the constraints on the decaying dark matter.

In this paper, we calculate the electromagnetic energy injection due to DM annihilation and the subsequent radiative processes based on the previously developed method~\cite{Kanzaki:2008qb,Kanzaki:2009hf}, and obtain constraints on the DM annihilation cross section from the CMB power spectrum.
We update the simulation code~\cite{Kanzaki:2008qb,Kanzaki:2009hf,Kawasaki:2015peu} by (1) including the helium interactions, (2) updating the  Planck data sets~\cite{Aghanim:2019ame},
(3) fixing the incorrect simplification on some interactions  
and (4) increasing the time steps from 50 bins to 200 bins for $z\in [0,10^4-1]$, which are summarized in section~\ref{sec_scat}.
We confirm that our simulation code reproduces the results from the Monte Carlo computation of the electron injection into the interstellar gas~\cite{shull1985x}.
The injected energy increases the ionization rate of the hydrogen and helium, and we modified the RECFAST code~\cite{Seager:1999km,Seager:1999bc,Wong:2007ym} to take account of the effect of extra ionization.
Then we can calculate the CMB power spectrum by the CAMB code \cite{Lewis:1999bs} with the modified RECFAST code.
Finally, we search for the allowed parameter space of the annihilation cross section and mass of DM using the CosmoMC code~\cite{Lewis:2002ah} with the Planck 2018 datasets~\cite{Aghanim:2019ame}.

{We introduce various radiative processes related to our calculation in section~\ref{sec_scat}. }
In section~\ref{sec_shower}, we show a formula which describes how the high-energy particles lose their energy in the early universe.
We then numerically calculate the fractions of the injected energy used for heating, ionization, and excitation.
In section~\ref{sec_InElec}, we discuss the modification of the RECFAST to include the effect of extra energy injection.
The results of the parameter search, using the CosmoMC code, are described in section~\ref{sec_Results}.

We use the following notation in this paper.
$n_\tx{HI}$ and $n_\tx{HII}$ describe the number densities of the neutral and ionized hydrogen.
The total number of the hydrogen atoms is $n_\tx{H}\equiv n_\tx{HI}+n_\tx{HII} $.
For helium, we define $n_\tx{HeI}$, $n_\tx{HeII}$, $n_\tx{HeIII}$ as number densities of neutral, partially ionized and fully ionized helium and $n_\tx{He}\equiv  n_\tx{HeI}+n_\tx{HeII}+n_\tx{HeIII}$.
Since we focus on the thermal history near and after the last scattering epoch, we approximate $n_\tx{HeIII}=0$.
We denote $n_e$ as the free electron number density.
The ionization fractions for hydrogen and helium are given by $x_\tx{H} \equiv n_\tx{HII}/n_\tx{H}$ and $x_\tx{He} \equiv n_\tx{HeII}/n_\tx{He}$, respectively.
The fraction of free electrons is given by 
$x_e \equiv n_e/n_\tx{H} =  x_\tx{H} + f_\tx{He} x_\tx{He}  $
with  the number fraction of helium atom to hydrogen atom $f_\tx{He} \equiv {n_\tx{He}}/{n_\tx{H} }  =  Y_{\rm He}/(4 (1 - Y_{\rm He})) \simeq 0.081$ and mass fraction $Y_{\rm He}\simeq 0.245$.

\section{Radiative processes}
\label{sec_scat}

In this section, we summarize newly added and updated processes in our calculation of electromagnetic shower.
The annihilation of DM injects high-energy particles, which immediately decay into photons, electrons, positrons, and neutrinos.
{The injected photon, electron and positron }lose their energy through various radiative processes. 
For energetic electrons, we consider the excitation and ionization of hydrogen and helium atoms, Coulomb scattering off the background electrons, the momentum transfer with neutral hydrogen,and inverse Compton scattering off the background photons. 
For energetic photons, we consider photoionization of the hydrogen and helium atoms, pair creation, Compton scattering off the background electrons, photon-photon scattering, and double-photon pair creation off the background photons.
{For energetic positrons, we consider the same scattering processes as those of electrons and the annihilation with the background electrons.}
The details of the above processes are given in~\cite{Kanzaki:2008qb}, and here we describe our improvements.

In the present work, we update our previous code by newly including the effect of helium.
We add the following new interactions :
 the excitation and collisional ionization of helium by high energy electrons, the photo-ionization of helium by high energy photons, and the pair creation of electrons and positrons induced by the photon-helium collisions.

We also update some naive assumptions in our previous code.
First, we used the common cross section for the  pair creation in nuclei in the previous code.
However, in the thermal plasma, the pair creation is induced by the free electrons, neutral and ionized hydrogen, and neutral and ionized of helium, and cross sections depend on those target particles.
We separately treat these pair creation processes.
Second, the cross section of the double-photon pair creation is sensitive to the energy of the background photons.
In the previous code, however, it was assumed that the background photons have a monochromatic energy distribution.
We include the energy distribution of the CMB photons to improve the calculation.
Third, the time resolution of the previous calculation was 50 log-spaced bins over the  $z\in [0,10^4-1]$, which was not sufficient.
In the revised calculation we take 200 bins for $z\in [0,10^4-1]$, which are enough to achieve the $\mathcal O(1)$\% precision of results.

We include the above modification in our calculation.
The explicit formulas of calculation are shown in the following, and we also comment on the treatment of positron, which is not explicitly explained in our previous work.

\subsection{Electron Interactions}
\label{sec_ElecHeli}

We include electron interactions with helium following Refs.~\cite{Slatyer:2009yq,Acharya:2019uba}.

For the excitation of the helium atom, we only include the excitation to 2p states.
For low energy electrons, we use the table of the cross section in Ref.~\cite{stone2002electron}.
For high energy electrons, we use the following formula~\cite{stone2002electron}:
\begin{align}
	\sigma(T)_\tx{He~exc} = \frac{4\pi a_0^2 \tx{Ry} }{T+B+E}
	\left[
		a\ln(T/\tx{Ry})
		+b+c\,\tx{Ry}/T
	\right]
	\frac{f_\mathrm{acc}}{f_\mathrm{sc}}
	,
\end{align}
with the Bohr radius $a_0$, the Rydberg constant Ry, the kinetic energy of incident electron $T$, the excitation energy $E=21.218~\eV$, and the binding energy $B=24.5874~\eV$. 
The dimensionless constants are given by $a=0.165601, ~b=-0.076942,~c=0.033306$ and  $(f_\mathrm{acc}/f_\mathrm{sc})=0.2762/0.2583$.
{
The final state electron has the energy $E_f = T-E$, and the differential cross section is given by $\frac{\df \sigma}{\df E_f}(T,E_f) = \sigma(T)_\tx{He~exc}~\delta(E_f-(T-E))$.
}

For the collisional ionization of the helium atom, we use the table of cross section in the CCC database \footnote{http://atom.curtin.edu.au/CCC-WWW/} for a low energy collision.
For a high energy collision, we use the following formula: \cite{arnaud1985updated,Slatyer:2009yq}
\begin{align}
	\sigma(E_i) = 10^{-14} \tx{cm}^{2}
	\frac{1}{u(I/\eV)^2}
	\left[
		A(1-u^{-1})
		+B(1-u^{-1})^2
		+C\ln u
		+Du^{-1} \ln u
	\right]
	,
\end{align}
with the ionization threshold $I$ and the normalized energy of incident electron $u=E_i/I$.
For the neutral helium, the constants are $A=17.8$, $B=-11.0$, $C=7.0$, $D=-23.2$, $I=24.6~\eV$.
{The differential cross section}
is approximately given by
\begin{align}
	\frac{\df\sigma(E,\epsilon)}{\df \epsilon} = \frac{A_{\sigma}(E)}{1+(\epsilon/\bar\epsilon)^2}
	\quad
	\tx{for}
	\quad
	0<\epsilon < \frac{1}{2}(E-I)
	,
\end{align}
where $E$ is the energy of a incident electron, $\epsilon$ is that of a ejected electron, $\bar \epsilon= 15.8~\eV$ and $A_{\sigma}(E)$ is a constant independent of $\epsilon$~\cite{opal1971measurements,1979ApJ...234..761S,Furlanetto_2010}.

\subsection{Photon Interactions }
\label{sec_PhotHeli}

Incident photons interact with hydrogen and helium atoms through the photo-ionization and the pair-creation processes. 
We obtain the helium photo-ionization cross section by interpolating the values in the table of~\cite{1979ApJS...40..815R}.

We update the cross section of pair creation based on Ref.~\cite{Slatyer:2009yq}.
We separately calculate the cross sections of pair creation with singly ionized hydrogen (HII) and helium (HeII)~\cite{Motz:1969ti}, neutral hydrogen (HI) and helium (HeI)~\cite{zdziarski1989absorption} and free electrons~\cite{RevModPhys.30.354}; they are given by
\begin{align}
    \sigma_{\rm HII}
    &=\sigma_{\rm HeII}
    =\alpha r_0^2
    \left(
        \frac{28}{9} \ln\left(\frac{2E_i}{ m_e} \right)
        -\frac{218}{27}
    \right),
    \\
    \sigma_{\rm HI}
    &=
    5.4\alpha r_0^2
    \ln\left(
        \frac{513E_i}{E_i +825 m_e}
    \right),
    \\
    \sigma_{\rm HeI}
    &=
     8.76\alpha r_0^2
    \ln\left(
        \frac{513E_i}{E_i +825 m_e}
    \right),
    \\
    \sigma_{{\rm free}\ e}
    &=
    \alpha r_0^2
    \left(
        \frac{28}{9} \ln\left(\frac{2E_i}{ m_e} \right)
        -\frac{100}{9}
    \right),
\end{align}
where $E_i$ is the initial energy of a photon, $\alpha$ is the fine structure constant, and $r_0= \alpha^2 a_0$ is the classical electron radius.
{The differential cross section} and the spectrum of produced particles is estimated by the Bethe-Heitler formula~\cite{Motz:1969ti}. {(See \cite{Kanzaki:2008qb} for the used formula in our code. )}

The incident photon may cause the double photon pair creation by interacting with the CMB photons once its energy overcomes the threshold $E_i \geq m_e^2/E_{\rm cmb}$ where $E_{\rm cmb}$ is the CMB photon energy.
We average the cross section over the energy distribution of the CMB photon to precisely estimate the effect of the double photon pair creation, following Ref.~\cite{Slatyer:2009yq}.

\subsection{Treatment of positron}
\label{sec_positron}

For the shower process, positrons are produced through the pair-creation and double-photon-pair-creation processes.
We assume that positrons lose their energy in the same way as electrons since both behave similarly for interactions at high energy~\cite{Slatyer:2009yq}. 
The produced positrons quickly lose their kinetic energy within one Hubble time since the background plasma is not transparent for positrons as well as electrons. 
Then, they annihilate the background electrons and produce photons with $E_\gamma= 511$~keV.
In our code, the kinetic energy loss processes of the positrons are treated in the same way as the electrons and the positron rest mass energy is converted to two photons with $E_\gamma= 511$~keV.

Both contributions are included in our shower code.

\section{Energy injection as shower process}
\label{sec_shower}

We briefly summarize the calculation of electromagnetic showers induced by DM annihilation based on \cite{Kanzaki:2008qb,Kanzaki:2009hf,Kawasaki:2015peu}.
We treat the energy injection as a perturbation on the standard thermal history of the $\Lambda$CDM model.
We focus on the redshift up to $z=10^4$, since the most important effects on the CMB come from the energy injection near and after the last scattering epoch.
In our calculation, we discretize the redshift and energy of particles and use logarithmic bins for them.

Let us describe energy injection and the induced electromagnetic shower in the thermal plasma.
Since the electromagnetic processes mainly produce high-energy photons, electrons, and positrons, we assume that the initial particle ``$a$'' is an electron or photon ($a=e$  or $\gamma$).
We consider an injected particle ``$a$" with energy $E_i$ at redshift $z=z_i$, which collides with another particle ``$b$", produces a cascade shower, and finally its energy is used for heating the background plasma, ionization, and excitation of the background hydrogen and helium.
Here we denote those processes by $\alpha=$ heat, ion$_\tx{H}$, ion$_\tx{He}$, exc, respectively.
We refer $\df\epsilon^a_\alpha(E_i,z_i,z_f)$ to the partial energy used for the process $\alpha$ during $[z_f,z_f-\df z_f]$.
Note that the redshift $z_f$ can be different from $z_i$ since the early universe can be transparent for the high-energy photons.
We want to evaluate the following quantity:
\begin{align}
    Q^a_\alpha (E_i,z_i,z_f) 
    = 
    \frac{\df\epsilon^a_\alpha(E_i,z_i,z_f) }{\df z_f} 
    .
\end{align}

When a particle is injected at slightly earlier time, $z_i \to z_i+\Delta z_i$, $Q^a_\alpha (E_i,z_i,z_f) $ is changed by scattering process during $\Delta z_i$.
Suppose that the injected particle ``$a$"  scatters off a particle ``$b$" with the cross section $\sigma(E_i)$, where the particle ``$b$" is a photon, electron, hydrogen, or helium in the background plasma.
The probability that the particle ``$a$'' loses its energy to $E_f$ for $\Delta z_i$ is given by 
\begin{align}
    \Delta p^a(E_i,E_f) 
    & =
    \frac{\df \nu^a (E_i,E_f)}{\df E_f} \frac{\df t}{\df z}\Delta z_i, 
\end{align}
where $\df \nu^a/\df E_f$ is the differential scattering frequency of the injected particle,
\begin{align}
     \frac{\df \nu^a (E_i,E_f)}{\df E_f}=
    n_t v_p \frac{\df \sigma^a(E_i,E_f)}{\df E_f}.
\end{align}
Here $n_t$ is the number of target particles, $\df \sigma^a/\df E_f$ is the differential cross section and $v_p$ is the relative velocity.
{The differential cross section is evaluated for the various radiative processes summarized in Section~\ref{sec_scat}.}
We denote $E^\tx{abs}_\alpha(E_i,E_f)$ as the partial energy consumed in the process $\alpha$, e.g. the binding energy of a hydrogen atom for the ionization process.
When the scattering process occurs at $z_i$,  $E^\tx{abs}_\alpha(E_i,E_f)$ contributes to $Q^a_\alpha (E_i,z_i,z_f)$ as $\Delta p^a (E_i, E_f) E^\tx{abs}_\alpha(E_i,E_f)\delta(z_i-z_f)$.
In the shower process, the multiple scatterings take place, and $Q^a_\alpha (E_i,z_i,z_f)$ depends on that of scattered particle, $Q^a_\alpha (E_f,z_i,z_f)$.
As a result, the $Q$ changes as
\begin{align}
     \Delta  Q^a_\alpha (E_i,z_i,z_f) 
    &=
    \delta(z_i-z_f) \int \df E_f~ \Delta p^a(E_i,E_f) E^\tx{abs}_\alpha
    \nonumber
    \\&+\int \df E_f ~
        \Delta p^a(E_i,E_f)
    \left[
         Q^a_\alpha (E_f,z_i,z_f) 
        +Q^b_\alpha (E_b,z_i,z_f) 
    \right]
    ,
    \label{eq_collisionterm}
\end{align}
where $E_b(E_i,E_f)(\equiv E_i-E_f - E_\alpha^{abs})$ is the final state energy of the particle $b$.

We rewrite $\Delta Q$ by the difference of $Q$ as
\begin{align}
     \Delta  Q^a_\alpha (E_i,z_i,z_f) 
    =& 
      Q^a_\alpha \left(
        E_i+
        \left[\tfrac{\df E_i}{\df z}\right]_{H}
        \Delta z_i
        ,z_i+\Delta z_i,z_f
      \right)
    -
     Q^a_\alpha (E_i,z_i,z_f)
     ,
     \label{eq_kineticterm}
\end{align}
where $\left[\tfrac{\df E_i}{\df z}\right]_{H}$ is the energy shift by the cosmic expansion and is given by $\left[\tfrac{\df E_i}{\df z}\right]_{H} = E_i/(z_i+1)$ for a relativistic particle and  $\left[\tfrac{\df E_i}{\df z}\right]_{H} = 2 E_i /(z_i+1)$ for a non-relativistic particle.
The injected energy at $z_i+\Delta z_i$ should be shifted to evaluate the difference due to radiative processes correctly.
Using Eqs.\eqref{eq_collisionterm} and \eqref{eq_kineticterm}, the time evolution of $Q$ is described by
\begin{align}
    \frac{\p  Q^a_\alpha (E_i,z_i,z_f)}{\p z_i} &
    +
    \left[\frac{\df E_i}{\df z}\right]_H\frac{\p  Q^a_\alpha (E_i,z_i,z_f)}{\p E_i}
    \nonumber
\\
    & = \delta(z_i-z_f) \int \df E_f~
    \frac{\df \nu^a (E_i,E_f)}{\df E_f}\frac{\df t(z_i)}{\df z}  E^\tx{abs}_\alpha
    \nonumber
    \\&+\int \df E_f ~
        \frac{\df \nu^a (E_i,E_f)}{\df E_f} \frac{\df t(z_i)}{\df z}
    \left[
         Q^a_\alpha (E_f,z_i,z_f) 
        +Q^b_\alpha (E_b,z_i,z_f) 
    \right]
    .
    \label{eq_BoltzRight}
\end{align}

In the numerical calculation, we discretize the energy range $E_i \in[1,10^{13}]\,\eV$ into $5200$ log-spaced bins and the redshift range $z_i,z_f\in [0,10^4-1]$ into $200$ log-spaced bins, respectively.
We solve Eq.\eqref{eq_differenceEq} based on Eq.\eqref{eq_BoltzRight} (see Appendix \ref{sec_numcalc} for details)
{
by using the radiative processes summarized in Section~\ref{sec_scat} and evaluate the $Q^a_\alpha(E_i,z_i,z_f)$ for wide ranges of injected energy, injected time, and consumed time. 
}

\begin{figure}
	\centering
	\includegraphics[width=.95\textwidth ]{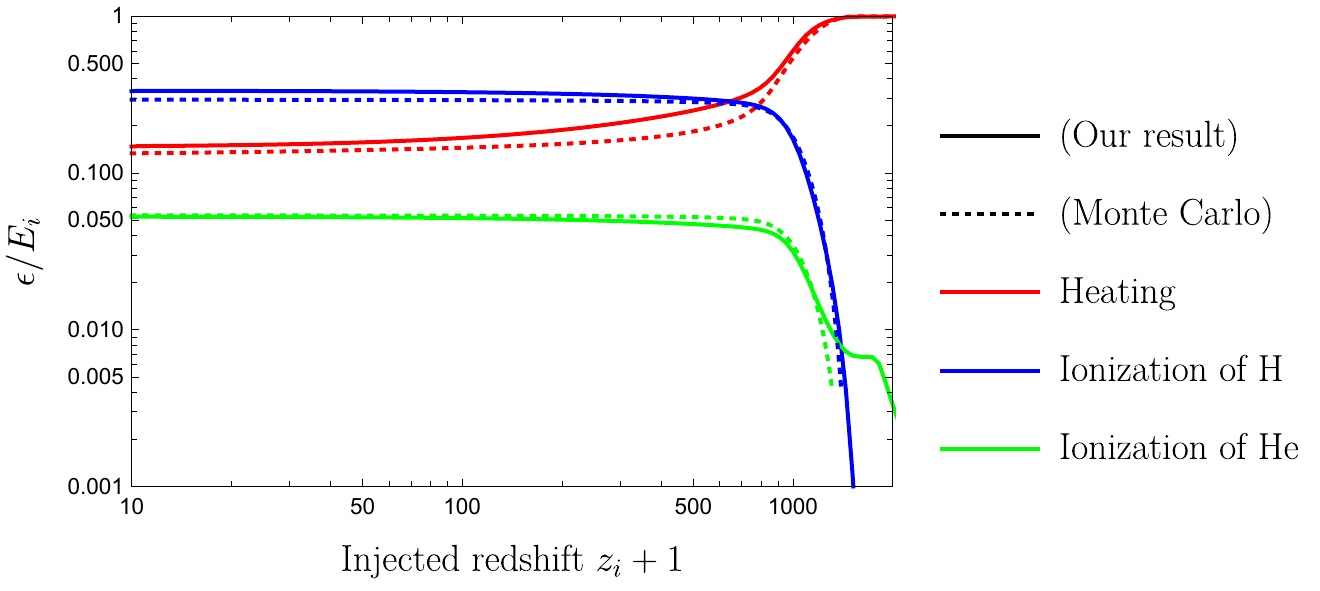}
	\caption{
    The energy ratios used for heating and ionization [Eq.~\eqref{eq_def_epsilon}] in case of the injected electron $E_i\simeq 3\,{\rm keV}$ and injected time $z_i$ integrated over $z_f\in [9,z_i]$
    {with $f_\mathrm{He}=0.1$.}
    The solid and dotted lines represent results of our simulation code and the Monte Carlo computation of the electron injection into the interstellar gas [Eqs.~\eqref{eq_ss_heat}-\eqref{eq_ss_Heion}] by Shull and Steenberg~\cite{shull1985x}, respectively.
    The red, blue, and green lines represent the energy ratio used for heating, ionization of hydrogen, and ionization of helium, respectively.
	}
	\label{fig_sum_chi_3keV}
\end{figure}

\begin{figure}[t]
	\centering
	\includegraphics[width=.8\textwidth ]{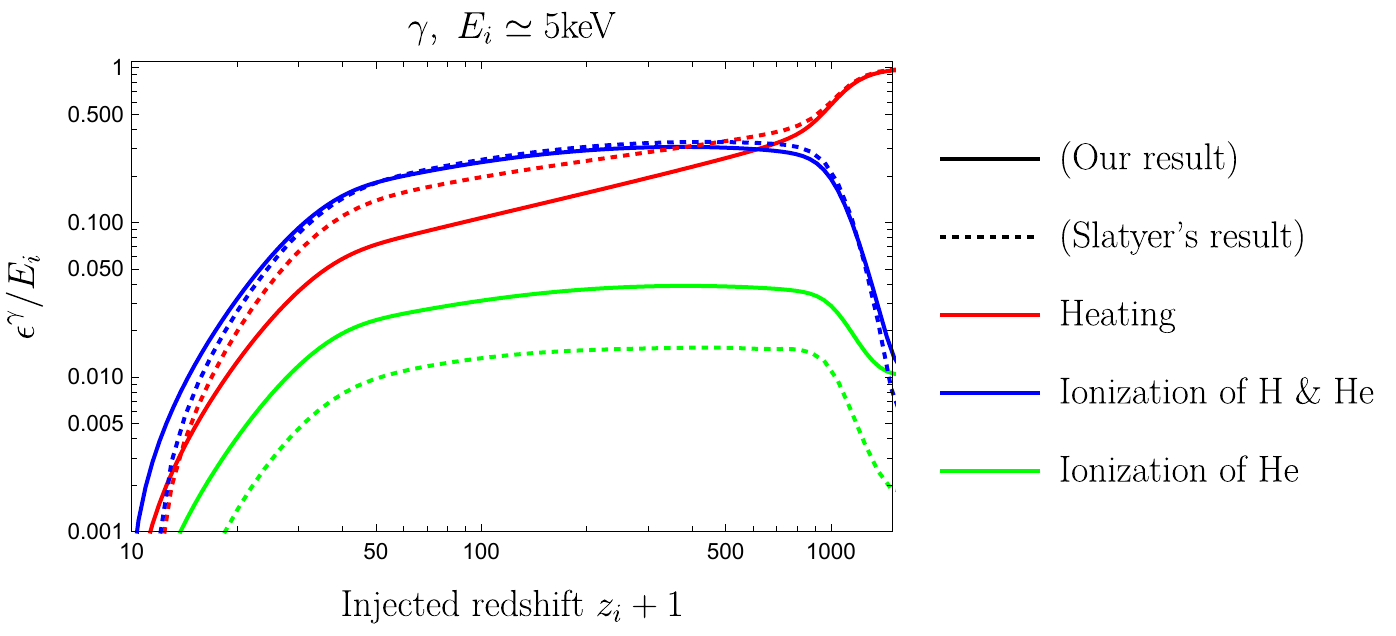}
	\includegraphics[width=.8\textwidth ]{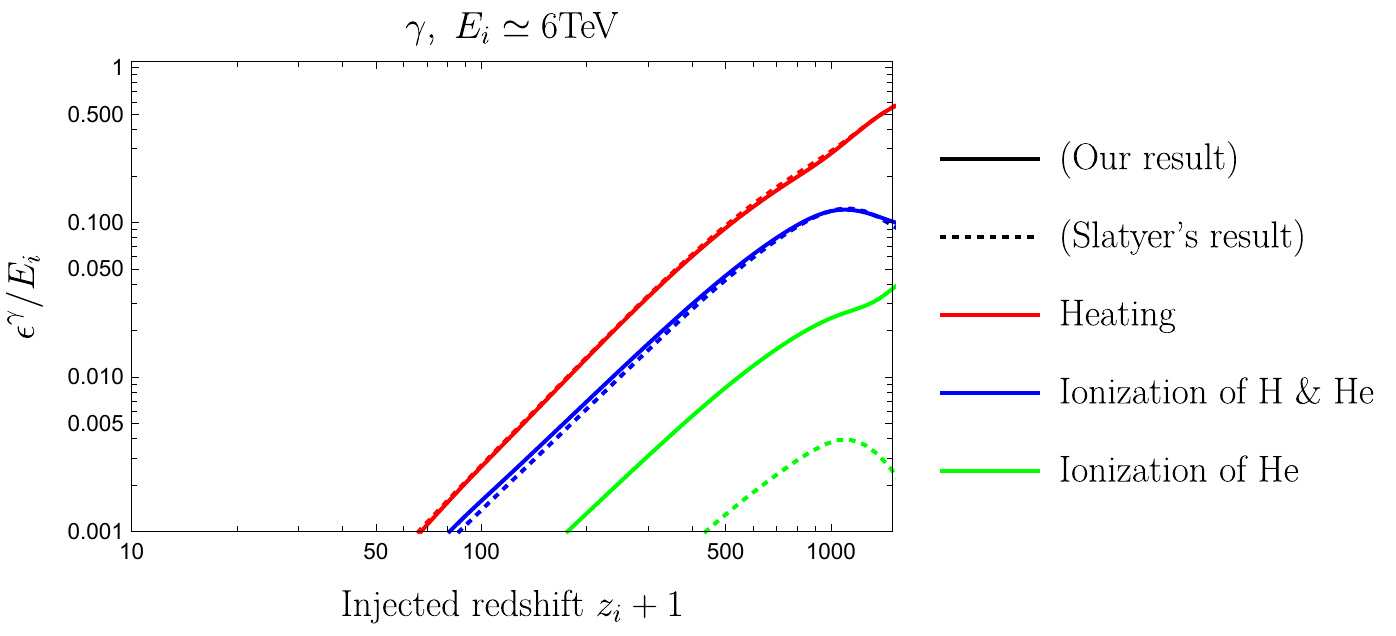}
	\caption{
    The energy ratio integrated over $z_f\in [9,z_i]$ for the injected photon with $E_i=5\,\keV$ and $6\,$TeV [Eq.~\eqref{eq_def_epsilon}].
    The red, blue, and green lines represent the energy ratios used for heating, the total ionization of hydrogen and helium, and ionization of only helium, respectively.
    The dotted lines are taken from Fig.~3 in~\cite{Slatyer:2015kla}, where we rename the sum of ``heating'' and ``sub-10.2\,eV continuum photons'' in~\cite{Slatyer:2015kla} to ``heating'' (red dotted line) since our code does not distinguish them.
	}
	\label{fig_sum_chi_comare_photon}
\end{figure}

\begin{figure}[t]
	\centering
	\includegraphics[width=.8\textwidth ]{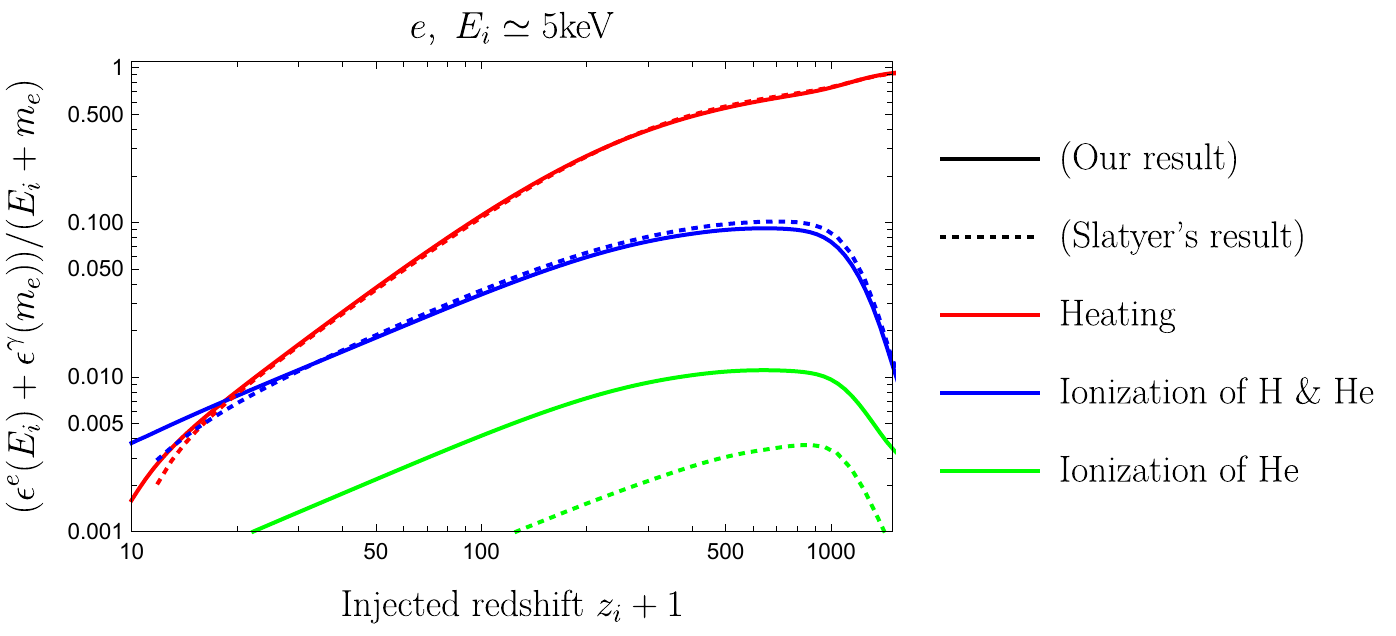}
	\includegraphics[width=.8\textwidth ]{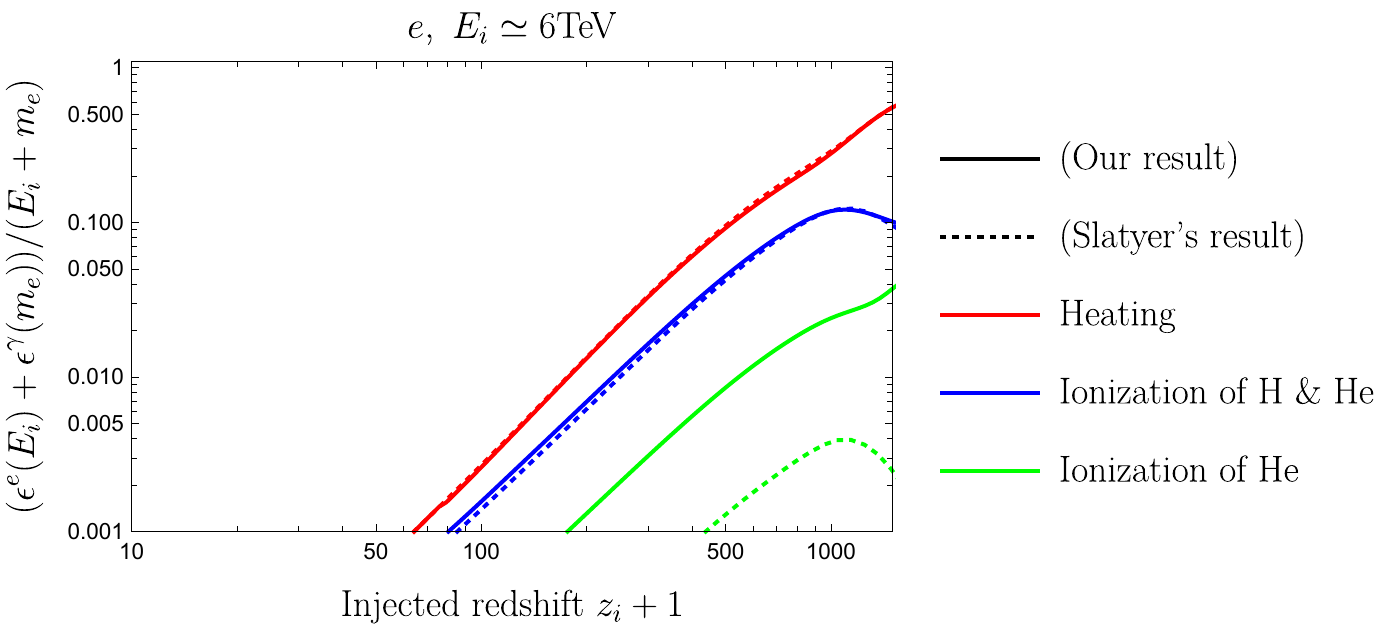}
	\caption{
	Same as Fig.~\ref{fig_sum_chi_comare_photon} except for the injected electron with its mass energy.
	}
	\label{fig_sum_chi_comare_electron}
\end{figure}

We check our calculation for low injected energy ($3$\,keV, $5$\,keV) and high injected energy ($6$\,TeV).
The results are shown in Figs.~\ref{fig_sum_chi_3keV}-\ref{fig_sum_chi_comare_electron}, where we plot the ratios of the initial energy to the energy used for heating and ionization during $z_f \in [9,z_i] $ as a function of the injected time $z_i$ for a given injected energy $E_i$, which is given by
\begin{align}
   \frac{\epsilon^a_\alpha(E_i,z_i)}{E_i} 
    \equiv 
    \int_{9}^{z_i}\df z_f
    \frac{Q^a_\alpha (E_i,z_i,z_f) }{E_i}
    .
    \label{eq_def_epsilon}
\end{align}
Except for a high energy photon, the $\epsilon^e_\alpha(E_i,z_i)/E_i$ is  mainly sourced by the $Q^a_\alpha (E_i,z_i,z_f)$ at $z_f\sim z_i$.
Following \cite{Slatyer:2015kla}, we choose the lower bound of the integration at $z_f =9$ to avoid the uncertainty of the cosmic reionization, which is neglected in our code.

In Fig.~\ref{fig_sum_chi_3keV}, we show $\epsilon^e_\alpha(E_i,z_i)/E_i$ for the injected electron with $E_i \simeq 3\,\keV$ (solid lines).
The red, blue, and green lines represent the energy ratios used for heating, ionization of hydrogen and helium, respectively.
{
We compare our results with the absorption rate calculated by Shull and Steenberg~\cite{shull1985x} to check whether our calculation properly includes the effect of helium.
 Shull and Steenberg performed the Monte Carlo computation of the electron injection into the interstellar gas, with two assumptions on the interstellar gas, $n_{\rm He}/n_{\tx H}=0.1$ and $x \equiv x_{\tx H}=x_{\rm He}$.
The derived fitting formula for the energy ratios to the injected electron energy $E_i\gg100\,\eV$ as a function of ionization fraction $x$ are given by
}
\begin{align}
    \frac{\epsilon^e_{\rm heat(SS)}}{E_i} 
    &=C\left[ 1- (1-x^{a})^{b}\right]
    \qcq &(C,a,b) = (0.9971,0.2663,1.3163)
    ,
    \label{eq_ss_heat}
    \\
    \frac{\epsilon^e_{\rm ionH(SS)}}{E_i} 
    &=C_{\rm H}(1-x^{a_{\rm H}})^{b_{\rm H}}
    \qcq &(C_{\rm H},a_{\rm H},b_{\rm H}) = (0.3908,0.4092,1.7592)
    ,
    \\
    \frac{\epsilon^e_{\rm ionHe(SS)}}{E_i} 
    &=C_{\rm He} (1-x^{a_{\rm He}})^{b_{\rm He}}
    \qcq &(C_{\rm He},a_{\rm He},b_{\rm He}) = (0.0554,0.4614,1.6660)
    .
    \label{eq_ss_Heion}
\end{align}
We plot Eqs.~\eqref{eq_ss_heat}-\eqref{eq_ss_Heion} by dotted lines in Fig.~\ref{fig_sum_chi_3keV} taking  $x=x_{\tx H}$. 
{
Here, we fix $n_{\rm He}/n_{\tx H}=0.1$ in our code to compare the effect of helium in the electromagnetic shower.
It is seen that our calculation well reproduces the ionization ratio of He in Ref.~\cite{shull1985x}.
Thus, we confirm that our code properly includes the helium interaction around $E_i\leq 3$\,keV.
}
We also comment on the slight difference of the heating and ionization ratios, where our results predict the larger ratios than those of Shull and Steenberg, which might be due to the different treatment of interactions.
{
A similar difference of the ionization ratio is reported in Fig.~1 of \cite{Slatyer:2015kla}. 
}
For $z\gtrsim 10^3$, our energy ratio used for the helium ionization is much larger than that in Ref.~\cite{shull1985x}, 
because the ionization fraction of helium is much smaller than that of hydrogen, i.e, $x_{\rm He} \ll x_{\rm H}$.

{
Next, we compare our results with that of the work by Slatyer~\cite{Slatyer:2015kla}, where the numerical table of results over $E_i\in [5\,\mathrm{keV}, 6\,\mathrm{TeV}]$ are provided.
The results at $E_i = 6\,\mathrm{TeV}$ include all information of interactions since the shower process at high energy includes the effects of the low energy processes.
Note that the interaction with highest energy scale is the double photon pair creation, which occurs at $E_i > m_e^2/E_\mathrm{cmb} \sim 1$\,TeV at $z=10^3$.
We also compare our results and the Slatyer's data at the lowest energy scale $5$\,keV to check the consistency of results on a low energy scale.
The energy absorption ratio for photon injection is shown in Fig.~\ref{fig_sum_chi_comare_photon}.
The dotted lines represent the results of the work by Slatyer~\cite{Slatyer:2015kla}.
The blue lines represent the total ionization of hydrogen and helium.
We consider the sum of the ``heating'' and ``sub-10.2\,eV continuum photons'' in Ref.~\cite{Slatyer:2015kla} as ``heating'' in this paper since our code does not distinguish them.
Compared to the photon injection, the definition of energy injection by an electron is different between our result and Slatyer's work.
Ref.~\cite{Slatyer:2015kla} considers the $e^+e^-$ pair injection and the total injected energy includes the rest mass energy of the electron and positron, so the energy absorption ratio is written by using our results as 
\begin{align}
    \frac{E_i}{E_i +m_e}
    \frac{\epsilon^e(E_i)}{E_i}
    +\frac{m_e}{E_i +m_e}
    \frac{\epsilon^\gamma(m_e)}{m_e},
\end{align}
where $E_i$ represents the kinetic energy of the electron and we assume that the rest mass energy of the positron is converted to two $511$\,keV photons.
We plot the energy absorption ratio of electron injection in Fig.~\ref{fig_sum_chi_comare_electron}.
}

One can see that our energy ratio used for the heating and total ionization (red and blue lines) are similar to those in Ref.~\cite{Slatyer:2015kla}.
On the other hand, our ratio used for the ionization of helium (green lines) is larger because of different treatment of the ionization process.
We separately treat ionization of hydrogen and helium, while Ref.~\cite{Slatyer:2015kla} uses the approximation that collisional ionization of hydrogen and helium are not distinguished for high injected energy $E_i>3$~keV.
The helium atoms are mostly ionized by the energetic photons with energy $E\sim(100-1000)\,\eV$.
Such photons are produced through inverse Compton scatterings by electrons with $E\gtrsim 10^7\,\eV$ at $z=10^3$ and lose their energy mainly by photo-ionization of helium.
Nevertheless, our calculation for the total energy used for ionization of hydrogen and helium atoms (blue line) is roughly the same as that in Ref.~\cite{Slatyer:2015kla}, and hence their effects on the CMB also do not change much.

We also mention the difference between our new calculation and the previous one~\cite{Kanzaki:2009hf}.
While the previous work assumes that all the baryon number is carried by the hydrogen, we correctly distribute the baryon number to hydrogen and helium.
Thus, the total number of bounded electrons is almost the same between the new and previous calculations when the helium and hydrogen are mostly neutral, which leads to a similar energy ratio for the total ionization.
The differences between our previous and new code are also induced by other modifications summarized in Sec.~\ref{sec_scat} and the updated background cosmological parameters.

\section{Effect of energy injection on CMB}
\label{sec_InElec}

In the previous section, we calculate the energy fraction $Q^a_\alpha (E_i,z,z')$ for $a=e$ and $\gamma$.
In general, DM may annihilate into various particles, e.g. $\mu^+\mu^-$ and $W^+W^-$, which finally decay into stable particles.
When DM has an annihilation channel $F$, we calculate the net energy injection by integrating the number of electrons, positrons, and photons of the final decay products as
\begin{align}
    Q^{(F)}_\alpha (E,z,z')
    &=
    \int \df E' 
    \bigg [
    \left( \frac{\df N_F^{(e^+)}(E,E') }{\df E'} +\frac{\df N_F^{(e^-)}(E,E') }{\df E'}  \right)
    Q^{(e)}_\alpha (E',z,z')
    \\ &+
    \frac{\df N_F^{(\gamma)}(E,E')  }{\df E'}
    Q^{(\gamma)}_\alpha (E',z,z')
    \bigg ],
\end{align}
where $dN_f^a(E,E')/dE'$ is the number of $a (=e^{\pm},\gamma)$ produced through the decay channel $F$. 
In this paper, we consider $F=e^{+}e^{-},~\gamma\gamma,~ \mu^{+}\mu^{-}$ and $W^{+}W^{-}$ as a typical decay channel.
(Notice that $dN_f^\gamma(E,E')/dE' = \delta(E-E')$ for $F=\gamma\gamma$ and $dN_f^{e^+}(E,E')/dE' =dN_f^{e^-}(E,E')/dE'= \delta(E-E')/2$ for $F=e^{+}e^{-}$.)
The decay fractions into $\mu^{+}\mu^{-}$ and $W^{+}W^{-}$ are calculated by the PYTHIA package~\cite{Sjostrand:2006za}.

We calculate the thermal history adding the following three terms in the RECFAST code:
\begin{align}
	-\left[ \frac{\df x_\tx{H}}{\df z} \right]_\tx{DM}
	&=
	\sum_F \int_z \frac{\df z'}{H(z')(1+z')}
	\frac{n_\tx{DM}^2(z')\braket{\sigma v}_F}{n_\tx{H}(z')}
	\frac{
	    Q_{\tx{ion}_\tx{H}}^{(F)}  (m_\tx{DM},z',z)
	}{B_\tx{H}}
	,
	\label{eq_xH}
\\
	-\left[ \frac{\df x_\tx{He}}{\df z} \right]_\tx{DM}
	&=
	\sum_F 
	\int_z
	\frac{\df z'}{H(z')(1+z')}
	\frac{n^2_{DM}(z') \braket{\sigma v}_F }{ n_\tx{He}(z') }
	\frac{
	    Q_{\tx{ion}_\tx{He}}^{(F)}  (m_\tx{DM},z',z)
	}{B_\tx{He}}
	,
	\label{eq_xHe}
\\
	-\left[  \frac{\df T_M}{\df z}  \right]	_\tx{DM}
	&=
	\sum_F 
	\int_z
	\frac{\df z'}{H(z')(1+z')}
	\frac{2}{3}
	\frac{n^2_{DM}(z') \braket{\sigma v}_F }{ (1+f_\tx{He}+x_e) n_{H}(z') }
	Q_{\tx{heat}}^{(F)}  (m_\tx{DM},z',z)
	,
	\label{eq_heat}
\end{align}
where $n_\text{DM}$ is the DM number density and the binding energies of HI and HeII are $B_\tx{H} = 13.6057$~eV and $B_\tx{He} = 24.5874$~eV.
For the fixed DM energy density $\rho_\tx{DM}$, the above modifications explicitly depends on $n_\tx{DM}^2\braket{\sigma v} m_\tx{DM}= \rho_\tx{DM}^2 \braket{\sigma v} /m_\tx{DM} $.
Thus, we expect the constraints on $\braket{\sigma v} /m_\tx{DM}$ by the CMB, which slightly depends on $m_\tx{DM}$ through $Q^{(F)}_\alpha(m_\tx{DM},z',z)/m_\tx{DM}$.

{
The modified thermal history changes the CMB power spectrum mainly through the enhanced Thomson scattering by the free electron.
The Thomson scattering suppresses the CMB temperature mode by increasing the optical depth, and at the same time, the CMB polarization mode is produced by the Thomson scattering. The latter effect mainly contributes to the constraint on the annihilation cross section.
This modification depends on the total number of ionization but not on the species of the ionized atom, e.g. hydrogen or helium~\cite{Galli:2013dna,Planck:2018vyg}.
}

\begin{figure}
	\centering
	\includegraphics[width=.8\textwidth ]{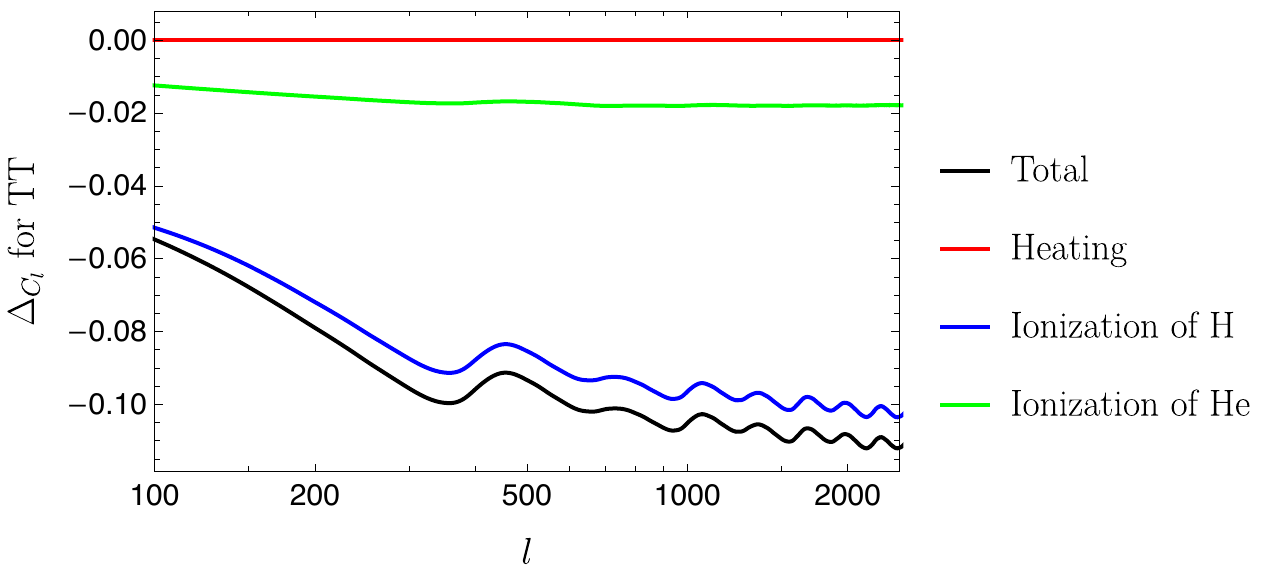}
	\includegraphics[width=.8\textwidth ]{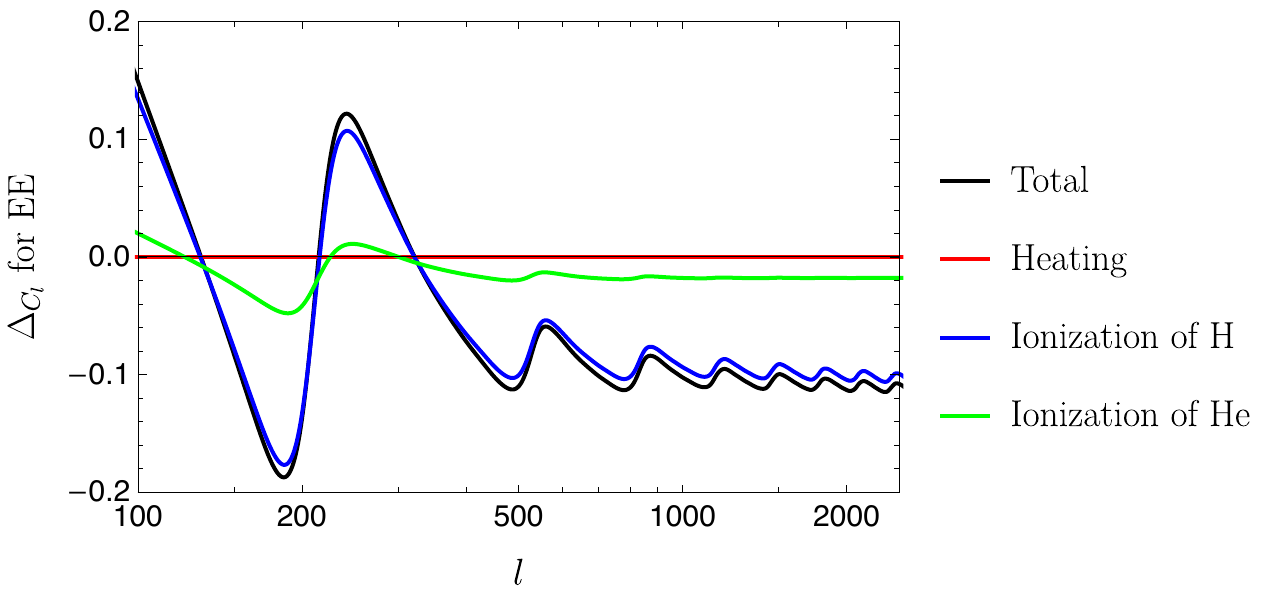}
	\caption{
    The difference of TT- and EE-mode power spectrum [Eq.~\eqref{eq_diff_CTT}] by the energy injection of DM annihilation into $e^+e^-$ channel for $m_{\rm DM} = 100$~GeV and $\braket{\sigma v}/m_{\rm DM} =1.0\times 10^{-26} {\rm cm}^3{\rm sec}^{-1}{\rm GeV}^{-1}$. 
    The blue, green, red and black lines represent the modified power spectrum including the effects of ionization of hydrogen in Eq.~\eqref{eq_xH}, ionization of helium in Eq.~\eqref{eq_xHe}, heating in Eq.~\eqref{eq_heat}, and all three contributions in Eqs.~\eqref{eq_xH}-\eqref{eq_heat}, respectively.
	}
	\label{fig_TTmode}
\end{figure}

In this paper, we calculate the CMB power spectrum using the CAMB~\cite{Lewis:1999bs} code with the modified RECFAST code~\cite{Seager:1999km,Seager:1999bc,Wong:2007ym}.
The energy injection changes the power spectrum of the CMB.
As an example, in Fig.~\ref{fig_TTmode} we calculate the difference of TT- and EE- mode power spectrum defined by
\begin{align}
    \Delta_{C_l}  = \frac{C^{\rm X}_{l}-C^{\rm X,(0)}_{l}}{C^{\rm X,(0)}_{l}}
    ,
    \label{eq_diff_CTT}
\end{align}
where $X= TT$ or $EE$, and $C^{\rm X,(0)}_{l}$ is the power spectrum of CMB without energy injection by DM and $C^{\rm X}_{l}$
is the spectrum with energy injection by $ e^+e^-$ channel for $m_{\rm DM} = 100$~GeV and $\braket{\sigma v}/m_{\rm DM} =1.0\times 10^{-26} {\rm cm}^3{\rm sec}^{-1}{\rm GeV}^{-1}$.
In Fig.~\ref{fig_TTmode}, the blue, green, red and black lines represent the modified power spectra including the effects of ionization of hydrogen [Eq.~\eqref{eq_xH}], ionization of helium [Eq.~\eqref{eq_xHe}], heating [Eq.~\eqref{eq_heat}], and all three effects, respectively.
It is seen that the dominant effect on the CMB power spectrum comes from the ionization of hydrogen, while the effect of helium is about 20\% of hydrogen and heating hardly affects the power spectrum.

\section{Constraints on the Dark matter annihilation}
\label{sec_Results}

In this section, we derive constraints on the DM annihilation cross section from the CMB observational data. 
We run the CosmoMC code~\cite{Lewis:2002ah} with the modified CAMB code as described in the previous sections and search for the DM parameter space $(m_\tx{DM}, \braket{\sigma v}/m_\tx{DM})$.
The range of DM mass is $m_\tx{DM}\in [1,10^4]\,\GeV$ for $2\gamma,~e^+e^-$, and $\mu^+\mu^-$ channel, and $m_\tx{DM}\in [80,10^4]\,\GeV$ for $W^+W^-$ channel.
The range of the cross section divided by mass is given by 
$ \braket{\sigma v}/m_\tx{DM}\in [0,3\times10^{-27}]\tx{ cm}^3\sec^{-1}\GeV^{-1} $ for $e^+e^-$ and $2\gamma$ channels,
 and  $ \braket{\sigma v}/m_\tx{DM}\in [0,6\times10^{-27}]\tx{ cm}^3\sec^{-1}\GeV^{-1} $ for $\mu^+\mu^-$ and $W^+W^-$  channels.
We use the top-hat prior for $\log(m_\tx{DM})$ and $\braket{\sigma v}/m_\tx{DM}$.
{
As for the cosmological model, we use the base-$\Lambda$CDM model in the Planck collaboration~\cite{Planck:2018vyg}. Except for the DM mass and annihilation cross section, we vary the six cosmological parameters $\{\Omega_b h^2,~\Omega_c h^2,~100\theta_{MC},~\tau,~ n_s,~{\rm{ln}}(10^{10} A_s)\}$ and neutrinos with normal mass hierarchy and $\sum m_\nu = 0.06$\,eV.
}

We use two combinations of data sets to constrain $\braket{\sigma v}/m_\tx{DM}$.
The first one is the CMB anisotropy from the Planck 2018 data set (``Planck only'') including low-l ($l\leq 29$) TT- and EE-mode (``lowl'' and ``simall\_EE'') , and high-l ($l\geq 30$)  TT-, TE- and EE-mode (``plik\_rd12\_HM\_v22\_TTTEEE'') \cite{Aghanim:2019ame}.
The other one, ``Planck+ext'', consists of the Planck CMB data and other cosmological data on galaxy correlation functions including the baryon acoustic oscillation (BAO) ~\cite{York:2000gk,Anderson:2013zyy,Ross:2014qpa,beutler20116df}, Dark Energy Survey (DES) ~\cite{Abbott:2017wau}, {and the CMB lensing power spectrum}.
{
The latter choice is expected to put stronger constraints than the former, since it can solve the parameter degeneracy.
}

\begin{figure}[t]
	\centering
\includegraphics[width=.95\textwidth ]{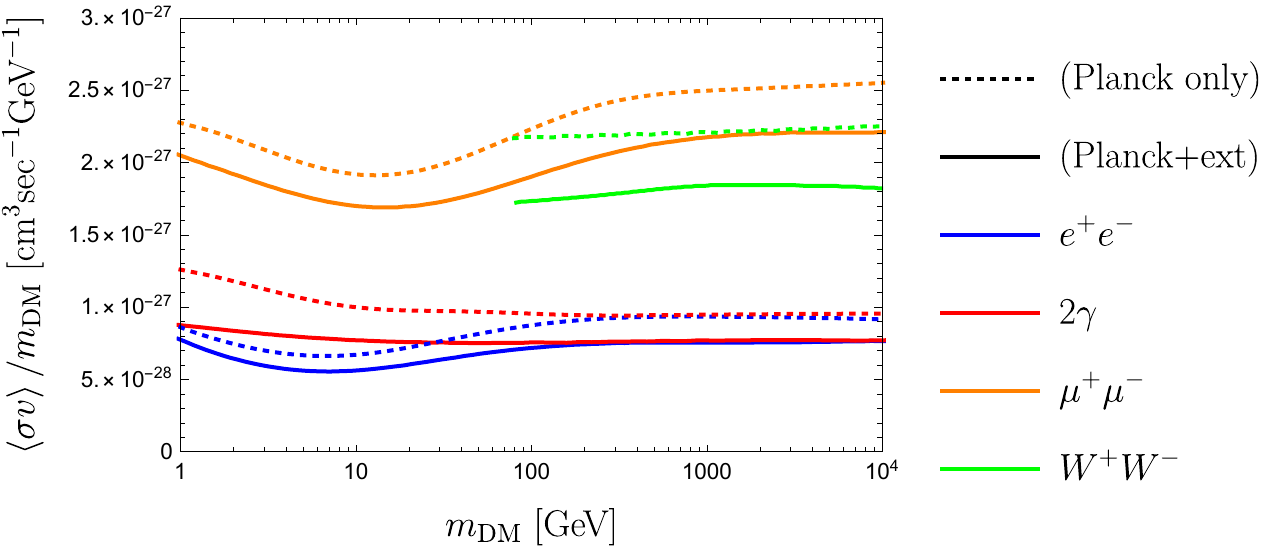}
	\caption{
		CMB constraints on the annihilation cross section for four decay channels of dark matter. 
		We plot the 95\% exclusion contours of $\braket{\sigma v}/m_{\rm DM}$ (vertical axes) for mass of dark matter (horizontal axes).
		We calculate the posterior distribution using two sets of data: ``Planck only'' uses the Planck CMB data for TT-, TE- and EE-modes~\cite{Aghanim:2019ame} and ``Planck+ext'' uses BAO~\cite{York:2000gk,Anderson:2013zyy,Ross:2014qpa,beutler20116df},  DES~\cite{Abbott:2017wau}, and the CMB lensing power spectrum data in addition to the Planck CMB data.
	}
	\label{fig_m_sigmam}
\end{figure}

\begin{figure}[h]
	\centering
	\includegraphics[width=.95\textwidth ]{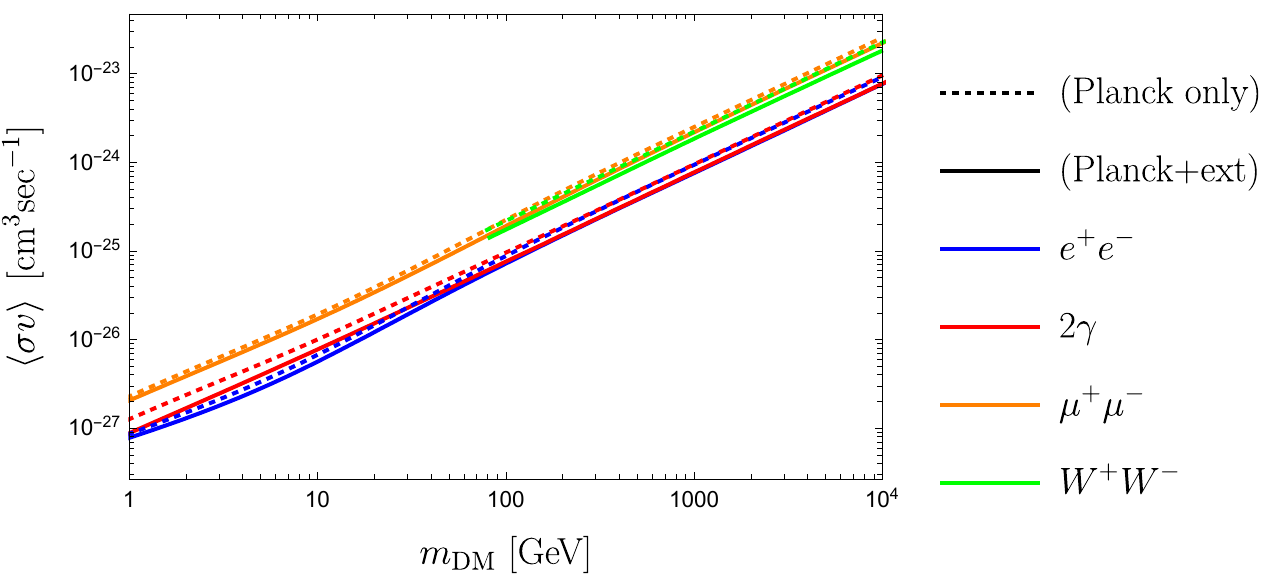}
	\caption{
		CMB constraints on the annihilation cross section for four DM annihilation channels. 
		We plot the 95\% exclusion contours of $\braket{\sigma v}$, which represent the same contours in Fig.~\ref{fig_m_sigmam}.
	}
	\label{fig_m_sigma}
\end{figure}

We show the constraints on the annihilation cross section in the  $(\braket{\sigma v}/m_{\rm DM},m_{\rm DM})$ plane in Fig.~\ref{fig_m_sigmam} and in the $(\braket{\sigma v}, m_{\rm DM})$ plane in Fig.~\ref{fig_m_sigma}, respectively.
The contours plotted by the GetDist code~\cite{Lewis:2019xzd} describe the $95\%$ exclusion region.
Comparing ``Planck only'' and ``Planck+ext'' in Fig.~\ref{fig_m_sigmam}, one can see that the extra data of BAO, DES {and CMB lensing power spectrum} tighten the constraints about $O(10\%)$.
{
The relative improvements are slightly dependent on the channel. It is not easy to figure out the physical reasons of this dependence because the role of the extra data (BAO, DES and CMB lensing) to constrain the DM model parameters is indirect and complicated. Note also that our numerical precision may be close to the limitation to resolve this slight channel dependence.
}
The constraints for $W^+W^-$ and $\mu^+\mu^-$ modes are about twice weaker than those for $e^+e^-$ and $2\gamma$ modes.
These features are consistent with our previous paper~\cite{Kawasaki:2015peu}. 
Since we change the prior of the Monte Carlo calculation of CosmoMC from the flat prior on $\braket{\sigma v}$ to the flat prior on $\braket{\sigma v}/m_{\rm DM}$, the upper bound on $\braket{\sigma v}$ is changed about $\mathcal O(10\%)$.

Constraints on the DM annihilation cross section has also been studied in many previous works~\cite{Padmanabhan:2005es,Kanzaki:2008qb,Kanzaki:2009hf,Slatyer:2009yq,Slatyer:2012yq,Galli:2013dna,Slatyer:2015jla,Slatyer:2015kla,Liu:2016cnk,Acharya:2019uba,Cang:2020exa}.
Among them, Refs.~\cite{Slatyer:2015jla,Slatyer:2015kla} include the effects of electromagnetic shower by injected particles based on Refs.~\cite{Slatyer:2009yq,Galli:2013dna}, where electromagnetic shower is calculated differently above and below the threshold energy $3$~keV. 
We confirm that our calculation reproduces similar results on the electromagnetic shower to their calculation in Sec.~\ref{sec_shower}. 
Our ``Planck only'' result for the $e^+e^-$ channel in Eq.~\eqref{fig_m_sigma} is consistent with Fig.~4 of Ref.~\cite{Slatyer:2015jla} within an accuracy of 10\%.

\section{Conclusion}
\label{sec_conclusion}

DM candidates like WIMPs are once in thermal equilibrium in the early universe and decouple from the thermal bath when their annihilation rate becomes slower than the cosmic expansion. 
The abundance of such thermal relics is determined by the annihilation cross section. 
Although the DM annihilation rate is smaller than the expansion rate, the annihilation of DM particles continues to take place after decoupling and may cause significant effects on cosmology and astrophysics.
We revisited the CMB constraints on the DM annihilation cross section.

In this paper, we calculated the electromagnetic energy injection due to DM annihilation and the subsequent radiative processes based on the previously developed method~\cite{Kanzaki:2008qb,Kanzaki:2009hf}, and obtain constraints on the annihilation cross section from the CMB power spectrum.
We have improved our previous calculation~\cite{Kawasaki:2015peu} on the following points: we have updated the observational data set including the Planck 2018 result~\cite{Aghanim:2019ame}, carefully included the helium interactions, fixed the incorrect simplification on some interactions, and improve the precision of each calculation. 
{
Although it was pointed out in \cite{Galli:2013dna,Planck:2018vyg} that the helium interaction in the shower process does not much change the constraints on DM annihilation cross section, the introduction of helium interaction improves the reliability of our code, and the consistent treatment of the shower process will become important for the other application of the energy injection into the thermal plasma.
}

Our calculation of the electromagnetic shower process is comparable to that in Ref.~\cite{Slatyer:2015jla} and our method is simpler.
While in Ref.~\cite{Slatyer:2015jla} the shower process is separately calculated by different codes for the low/high energy region, our code can consistently calculate the shower process in both high and low energy regions without unnecessary assumptions.
We use the recursive method by separating the ``discrete'' and ``continuous'' interactions, where the scattered particle loses larger (smaller) energy than the size of energy bin $\Delta E$ in discrete (continuous) interactions.
Our results are fairly consistent with the results of Ref.~\cite{Slatyer:2015jla} for the $e^+e^-$ channel.
We also confirm that our calculation reproduces the Monte Carlo calculation on the electromagnetic shower of particle scattering given in Ref.~\cite{shull1985x}.
Our results are summarized in Fig.~\ref{fig_m_sigma}, where the 95\% exclusion limits are shown for four decay channels: $2\gamma$, $e^+e^-$, $\mu^+\mu^-$ and $W^+W^-$.

\appendix

\section{Numerical scheme of shower code }
\label{sec_numcalc}

In this Appendix, we describe our scheme of the numerical calculation.
We discretize the energy and redshift in log-scale.
First, we define the normalized energy ratio $\chi^a_\alpha$ as
\begin{equation}
    \chi^a_\alpha(E_i,z_m,z_n) \equiv \Delta z_n~ Q^a_\alpha (E_i,z_m,z_n)/E_i,
\end{equation}
with a size of bin $\Delta z_n$.
Here $z_m$ and $z_n$ represent the redshifts at injection and absorption, respectively.
Using $\chi_\alpha^a$ we rewrite Eq.\eqref{eq_kineticterm} as
\begin{align}
    \Delta  Q^a_\alpha (E_i,z_m,z_n)
     & = 
    \frac{E_i}{\Delta z_n}
    \left[
     \chi^a_\alpha (E_i,z_m,z_n)
     -
       \frac{E_i-\left[\tfrac{\df E_i}{\df z}\right]_{H}\Delta z_m}{E_i}
        \chi^a_\alpha \left(
        E_i-
        \left[\tfrac{\df E_i}{\df z}\right]_{H}
        \Delta z_m
        ,z_m-\Delta z_m,z_n
      \right)
    \right]
    ,
    \label{eq_BolzDiscleft}
\end{align}
where we slightly change the time variable from $z_i$ to $z_m-\Delta z_m$.
In the same way, we rewrite Eq.\eqref{eq_collisionterm} as
\begin{align}
     \Delta  Q^a_\alpha (E_i,z_m,z_n) 
    &=
    \frac{E_i}{\Delta z_n}
    \bigg[
    \Delta z_n \delta(z_n-z_m) \sum_j \Delta E_j~ \Delta p^a(E_i,E_j) 
    \frac{E^\tx{abs}_\alpha}{E_i}
    \nonumber
\\&+
    \sum_j \Delta E_j~
       \Delta p^a(E_i,E_j)
    \left(
         \frac{ E_j}{E_i} \chi^a_\alpha (E_j,z_m,z_n) 
        +\frac{E_b}{E_i} \chi^b_\alpha (E_b,z_m,z_n) 
    \right)
    \bigg]
\\
 &=
    \frac{E_i}{\Delta z_n}
    \bigg[
    \delta_{m,n} \sum_j \Delta P^a(E_i,E_j,z_m)  
    \frac{E^\tx{abs}_\alpha}{E_i}
    \nonumber
\\&+
    \sum_j \Delta P^a(E_i,E_j,z_m)  
    \left(
         \frac{ E_j}{E_i} \chi^a_\alpha (E_j,z_m,z_n) 
        +\frac{E_b}{E_i} \chi^b_\alpha (E_b,z_m,z_n) 
    \right)
    \bigg]
    ,
\end{align}
where we use $\Delta z_n \delta(z_n-z_m)\to \delta_{m,n}$ and define the differential scattering probability as
\begin{align}
   \Delta P^a(E_i,E_j,z_m) 
   &\equiv
     \Delta\nu^a (E_i,E_j) \frac{\df t(z_i)}{\df z} \Delta z_n
    ,
    \label{eq_DelPNaive}
\end{align} 
with the differential collisional frequency:
\begin{align}
    \Delta \nu^a (E_i,E_j)
   & \equiv
    \Delta E_j~ \frac{\df \nu^a (E_i,E_j)}{\df E_j}
    = n_t v_p \frac{\df \sigma^a (E_i,E_j)}{\df E_j} \Delta E_j 
    .
    \label{eq_partcollfreq}
\end{align}
 
We should be careful when the difference of energy before and after scattering is small.
At first, when some fraction of the scattered particles belong to the same energy bin as the initial particles do, the effective scattering probability should be smaller than that calculated by Eq.~\eqref{eq_partcollfreq}. 
We call such scattering ``continuous loss", and its collisional frequency is given by the ratio of losing energy to $\Delta E_i$ as
\begin{align}
    \Delta\nu^a(E_i,E_{i-1}) =
    &  \frac{1}{\Delta E_i} \left[ \frac{\df E}{\df t}  \right]_c
    ,
\end{align}
where $\left[ \tfrac{\df E}{\df t}  \right]_c$ is the energy loss by the continuous loss process.
Second, when the plasma is optically thick for an injected particle, the sum of the scattering probability during $\Delta z_m$ can exceed $1$. 
In this case, the scattering probability at one channel is given by the ratio of its differential scattering frequency to the total as
\begin{align}
    \Delta P^a(E_i,E_j,z_m) &= 
    \frac
    { \Delta\nu^a(E_i,E_j)}
    {\sum_k \Delta\nu^a(E_i,E_k) +\nu_H^a(E_i) }.
    \label{eq_DelPDiscrete}
\end{align}
Here $\nu_H^a(E_i)$ is the effective frequency of the Hubble friction defined by
\footnote{
Note that Eq.\eqref{eq_DelPDiscrete} coincides with Eq.\eqref{eq_DelPNaive} in the continuous limit of time step since we can derive 
\begin{align}
    \nu_H (E_i) &= \frac{H(z)E_i}{\Delta z_m \frac{\df E_i}{\df z}}   =\frac{H(z)E_i}{\Delta t \frac{\df E_i}{\df t}}
    =\frac{H(z)E_i}{\Delta t H E_i}
    =(\Delta t)^{-1}.
\end{align}
}
\begin{align}
	\nu_H (E_i) \equiv \frac{H(z)E_i}{\Delta_z E_i}  ,
\end{align}
where $\Delta_z E_i = E_i\, \Delta z_m/(1+z_m) $.
We also normalize the ratio of redshifted particles in the second term of Eq.\eqref{eq_BolzDiscleft} using the following factor:
\begin{align}
     \Delta P^a_H(E_i,E_j,z_m) &= 
    \frac
    { \nu_H^a(E_i)  }
    {\sum_k \Delta\nu^a(E_i,E_k) +\nu_H^a(E_i) }.
\end{align}

Combining these formulas, we can perform the numerical calculation of the shower process based on the following formula:
\begin{align}
	\chi^a_{\alpha}(E_i,z_m,z_n) &=
	\nonumber
    \delta_{mn}
    \sum_j \Delta P^a(E_i,E_j,z_m)  
    \frac{E^\tx{abs}_\alpha}{E_i}
	\\&\nonumber
	\quad +\sum_{j<i}	\Delta P^a(E_i,E_j,z_m)
	\left[
        \frac{ E_j}{E_i} \chi^a_\alpha (E_j,z_m,z_n) 
        +\frac{E_b}{E_i} \chi^b_\alpha (E_b,z_m,z_n) 
	\right]
	\\&
	\quad 
	+\delta_{m\neq n}
	P_H^a(E_i,z_m) \chi^a_\alpha (E_i- \Delta_z E_i,  z_{m-1}, z_n ) \frac{E_i-\Delta_z E_i}{E_i},
	\label{eq_differenceEq}
\end{align}
where $\delta_{m\neq n}$ is 0 for $m=n$ and  1 for $m\neq n$.

To check the energy conservation, we also calculate the energy used for the cosmic expansion (we denote this as process ``$z$'', i.e. $\alpha=z$).
Then we can write $\chi^a_z$ as
\begin{align}
	\chi_z^a(E_i,z_m,z_n) &=
	\nonumber
    \delta_{mn}
    P_H^a(E_i,z_m)
    \frac{\left[\tfrac{\df E_i}{\df z}\right]_{H}
        \Delta z_m}{E_i}
	\\&\nonumber
	\quad +\sum_{j<i}	\Delta P^a(E_i,E_j,z_m)
	\left[
        \frac{ E_j}{E_i} \chi^a_z (E_j,z_m,z_n) 
        +\frac{E_b}{E_i} \chi^b_z (E_b,z_m,z_n) 
	\right]
	\\&
	\quad 
	+\delta_{m\neq n}
	P_H^a(E_i,z_m) \chi^a_z (E_i- \Delta_z E_i,  z_{m-1}, z_n ) \frac{E_i-\Delta_z E_i}{E_i}.
\end{align}
We also calculate the energy fraction $\chi^a_r$ which is not used for any process until $z_m$ and $\chi^a_r$ is written as
\begin{align}
    \chi^a_r(E_i,z_m,z_n) &=
	\nonumber
    \delta_{mn}
    P_H^a(E_i,z_m)
    \frac{E_i - \left[\tfrac{\df E_i}{\df z}\right]_{H}
        \Delta z_m}{E_i}
	\\&\nonumber
	\quad +\sum_{j<i}	\Delta P^a(E_i,E_j,z_m)
	\left[
        \frac{ E_j}{E_i} \chi^a_r (E_j,z_m,z_n) 
        +\frac{E_b}{E_i} \chi^b_r (E_b,z_m,z_n) 
	\right]
	\\&
	\quad 
	+\delta_{m\neq n}
	P_H^a(E_i,z_m) \chi^a_r (E_i- \Delta_z E_i,  z_{m-1}, z_n ) \frac{E_i-\Delta_z E_i}{E_i}.
\end{align}

We then calculate $\chi^a_\alpha(E_i,z_m,z_n)$ ($\alpha=$ heat, ion$_\tx{H}$, ion$_\tx{He}$, exc, z, r) from low to high energy and from low to high redshift.
We have confirmed the energy conservation, $\sum_\alpha \chi_\alpha^a(E_i,z_m,z_m) =1$ for $z_m=z_n$ and $\sum_\alpha \chi_\alpha^a(E_i,z_m,z_n) =\chi_r^a(E_i,z_m,z_{n+1}) $ for $z_m>z_n$.

\begin{acknowledgments}
This work was supported by JSPS KAKENHI Grant Nos. 17H01131 (M.K.), 17K05434 (M.K.), 20H05851(M.K.), 21K03567(M.K.), JP19J21974 (H.N.), 18K03609 (K.N.), 17H06359 (K.N.), 
Advanced Leading Graduate Course for Photon Science (H.N.), and 
World Premier International Research Center
Initiative (WPI Initiative), MEXT, Japan (M.K.).
\end{acknowledgments}

\small
\bibliographystyle{apsrev4-1}
\bibliography{Ref_EnergyInject.bib}
\end{document}